\documentclass[12pt]{article}

\title{How can one find nonlocality in Bohmian mechanics?}
\author{Andrei Khrennikov\footnote{Supported by the EU Human Potential Programme under 
contract N. HPRN-CT-2002-00279}\\
International Center for Mathematical\\
Modeling in Physics and Cognitive Sciences,\\
MSI, University of V\"axj\"o, S-35195, Sweden\\
Email:Andrei.Khrennikov@msi.vxu.se}
\begin{document}
\maketitle

\begin{abstract}We perform the probabilistic analysis of the pilot wave formalism. 
From the probabilistic point of view it is not so natural 
to follow D. Bohm and to consider nonlocal interactions between parts of (e.g.)
a two-particle system. It is more natural to consider dependence of
corresponding preparation procedures and the propagation of the 
initial correlations between preparation procedures. In the pilot wave formalism
it is more natural to speak about  correlations of the initial pilot 
waves which propagate with time.
\end{abstract}

{\bf 1. Introduction}

Traditionally the pilot-wave quantum formalism, Bohmian mechanics (see, e.g.,  [1]-[3]), 
is considered as a 
{\it{nonlocal model,}} see, e.g., D. Bohm and B. Hiley [2], p. 58:
{\scriptsize{``In our interpretation of quantum theory, we see that the interaction of 
parts is determined by something that cannot be described solely in terms of
there parts and their preassigned interrelationships.''}}
\footnote{It is interesting to notice that J. Bell was 
very excited by the Bohmian nonlocality. This was one of the main sources 
of the Bell's inequality. I also understand well that people belonging to the Bohmian 
community have various views to Bohmian nonlocality and these views are in the
process of  permanent  changing, see, e.g. [4]. Here I present just the common view 
to the Bohmian mechanics as a nonlocal theory.}

In this note we analyse the pilot wave model and Bohm's arguments about 
nonlocality from the probabilistic viewpoint. Our analysis demonstrated that
Bohm's nonlocal interpretation of many-particles systems is not so natural. 
Instead of a nonlocal interaction, it would be more natural to consider 
dependence produced by the preparation procedure for the initial state 
and propagation of this dependence in time (with modifications owed to interactions of subsystems).

{\bf 2. Nonlocality or dependence?}

Let us consider two particle systems in which particles do not interact.
This model can be described by  the Hamiltonian:
$$
\hat H= \hat H_1 + \hat H_2 =(\frac{-h^2}{2 m_1} \Delta_1 + V_1) 
+ (\frac{-h^2}{2 m_2} \Delta_2 + V_2),
$$
where $\Delta_j$ are Laplacians and $V_j= V_j (x_j), j=1,2,$ are potentials.

The important thing is that the operators $\hat H_1$ and $\hat H_2$ commute, $[\hat H_1, \hat H_2]=0.$ Thus:
\begin{equation}
\label{E}
e^{\frac{-it\hat H}{h}}=e^{\frac{-it \hat H_1}{h}} e^{\frac{-it \hat H_2}{h}}
\end{equation}

Let us now consider the corresponding Schr\"odinger equation:
\[i h \frac{\partial \psi}{\partial t} (t, x_1, x_2)=\hat H \psi (t, x_1, x_2)\]
\[\psi (0, x_1, x_2)=\psi_0 (x_1, x_2).\]
Suppose now that at the initial instant of time particles composing a two particle
system were prepared independently\footnote{Well there always can exist some correlations
in preceding histories of systems, but we assume that  it is possible to neglect all those correlations.}
\begin{equation}
\label{IN}
\psi_0 (x_1, x_2)=\varphi_1 (x_1) \varphi_2{(x_2}).
\end{equation}
This condition implies that
\begin{equation}
\label{IN2}
{\bf{P}}_{\psi_0} (x_1, x_2)={\bf P}_{\varphi_1} (x_1) {\bf P}_{\varphi_2} (x_2).
\end{equation}
Thus the probability to find parts of a two-particle system having the 
state $\psi$ in points $x_1$ and $x_2$, respectively, is factorized into probabilities for 
the position observations on particles in the states $\varphi_1$ and $\varphi_2,$ respectively.

We remark that the factorization condition $(\ref{IN})$ for a wave
function also implies the factorization condition in the momentum representation (and vice versa):
\begin{equation}
\label{IN3}
\psi_0 (p_1, p_2) = \varphi_1 (p_1) \varphi_2 (p_2),
\end{equation}
where
$$
\psi_0 (p_1, p_2) = \int e^{i p_1 x_1 + i p_2 x_2} \psi_0 (x_1, x_2) dx_1 dx_2.
$$ 
Thus the condition  (\ref{IN})  also implies the factorization of the probability for the 
momentum measurements:
\begin{equation}
\label{IN4}
{\bf P}_{\psi_0} (p_1, p_2)={\bf P}_{\varphi_1} (p_1), {\bf P}_{\varphi_2}(p_2)
\end{equation}
Hence under the condition (\ref{IN}) preparations are independent both with respect to the position and momentum.

I did not perform detailed analysis, 
but it seems that conditions (\ref{IN2}) and, (\ref{IN4}) together imply the condition (\ref{IN})
(and hence (\ref{IN3})). Thus by considering the factorization (\ref{IN}) of the wave function 
we, in fact, consider preparations which are independent with 
respect to both fundamental variables, the position and the momentum.

By (\ref{E}) we have 
$$
\psi(t, x_1, x_2)=e^{\frac{-it \hat H}{h}} \psi_0 (x_1, x_2)=\varphi_1 (t, x_1) \varphi_2(t, x_2), 
$$
where $\varphi_1(t)$ and $\varphi_2(t)$
are solutions of the Schr\"odinger equations with Hamiltonians $\hat H_1$ and $\hat H_2$ 
and the initial conditions $\varphi_1$ and $\varphi_2.$

Thus independence (both in the position and momentum representations)
is preserved in the process of evolution. We see that if there were
no correlations between preparations of statistical ensembles of particles composing 
two-particle systems then (in the absence of interactions between particles, i.e., 
$\hat V=\hat V_1 + \hat V_2, [\hat V_1, \hat V_2]=0)$ there is no any trace of the 
Bohmian nonlocality.

Even if we use the individual interpretation of a wave function and associate a wave
function with a single (e.g. two particle) system as a pilot wave, then in the light of the
presented probabilistic analysis it is more natural to consider correlation between initial
pilot waves (which propagate according to the Schr\"odinger equation) and not a nonlocal interaction
in the process of evolution.
Not particles are "guided in a correlated way", cf. [2], p. 57, but initially correlated
waves propagate according to the Schr\"odinger equation.

Of course, if the initial state cannot be factorized, i.e., the 
preparation of the initial ensemble of two-particle systems could not be 
split into two independent preparation procedures (both with respect to the position and the momentum variables), 
then the initial dependence will propagate with time.

If we use the language of the Bohmian mechanics we can say 
if initially pilot waves corresponding to parts of two-particle systems 
were correlated their correlation would not disappear immediately and 
it will propagate in time (by the law given by the Schr\"odinger equation).

{\bf{Conclusion.}}
The pilot wave formalism need not be interpreted in the
Bohmian way\footnote{As I know from some private conversations, L. De Broglie personally did not support
the nonlocal interpretation of the pilot wave approach (unfortunately I do not have
a precise reference).} -- as a formalism where ``interactions can therefore be described as {\it{nonlocal.}''}

It is natural to interpret this formalism by considering
propagation of correlations induced by state preparations.

I would like to thank B. Hiley and D. D\"urr for a discussion on nonlocality of Bohmian mechanics.
Especially I am thankful to them for the explanation that (despite a rather common opinion)
indistinguishability of particles cannot be 
considered as a fundamental feature of quantum multi-particle systems which induces "nonclassical probabilistic
behaviour".

{\bf References}

1.  D. Bohm, {\it Quantum theory, Prentice-Hall.} 
Englewood Cliffs, New-Jersey, 1951.

2. D. Bohm and B. Hiley, {\it The undivided universe:
an ontological interpretation of quantum mechanics.}
Routledge and Kegan Paul,  London, 1993.

3. P. Holland, {\it The quantum theory of motion.} Cambridge University press, Cambridge, 1993.

4. B. Hiley, From the Heisenberg picture to Bohm: a new perspective on active information and its
relation to Shannon information.
Proc. Int. Conf. {\it Quantum Theory: Reconsideration
of Foundations.} Ed. A. Yu. Khrennikov, Ser. Math. Modelling, vol 2.,
141-162, V\"axj\"o Univ. Press, 2002, see http://www.msi.vxu.se/forskn/quantum.pdf
\end{document}